# SOLID STATE QUANTUM BIT CIRCUITS

Daniel Esteve and Denis Vion

*Quantronics, SPEC, CEA-Saclay,*
*91191 Gif sur Yvette, France*

# Contents

## 1. Why solid state quantum bits?

Solid state quantum bit circuits are a new type of electronic circuit that aim to implement the building blocks of quantum computing processors, namely the quantum bits or qubits. Quantum computing [1] is a breakthrough in the field of information processing because quantum algorithms could solve some mathematical tasks presently considered as intractable, such as the factorisation of large numbers, exponentially faster than classical algorithms operated on sequential Von Neumann computers. Among the various implementations envisioned, solid state circuits have attracted a large interest because they are considered as more versatile and more easily scalable than qubits based on atoms or ions, despite worse quantumness. The 2003 Les Houches School devoted to *Quantum Coherence and Information Processing* [2] has covered many aspects of quantum computing [3], including solid state qubits [4–7]. Superconducting circuits were in particular thoroughly discussed. Our aim is to provide in this course a rational presentation of all solid state qubits. The course is organised as follows: we first introduce the basic concepts underlying quantum bit circuits. We classify the solid state systems considered for implementing quantum bits, starting with semiconductor circuits, in which a qubit is encoded in the quantum state of a single particle. We then discuss superconducting circuits, in which a qubit is encoded in the quantum state of the whole circuit. We detail the case of the quantronium circuit that exemplifies the quest for quantum coherence.

### 1.1. From quantum mechanics to quantum machines

Quantum Computing opened a new field in quantum mechanics, that of quantum machines, and a little bit of history is useful at this point. In his seminal work, Max Planck proved that the quantisation of energy exchanges between matter and the radiation field yields a black-body radiation law free from the divergence previously found in classical treatments, and in good agreement with experiment. This success led to a complete revision of the concepts of physics. It took nevertheless about fifty years to tie together the new rules of physics in what is now called quantum mechanics. The most widely accepted interpretation of quantum mechanics was elaborated by a group physicists around Niels Bohr in Copenhagen. Whereas classical physics is based on Newtonian mechanics for the dynamics of any system, and on fields, such as the electromagnetic field described by Maxwell's equations, quantum mechanics is based on the evolution of a system inside a Hilbert space associated to all its physically possible states. For example, localised states at all points in a box form a natural basis for the Hilbert space of a particle confined in this box. Any superposition of the basis states is a possible physical state. The evolution inside this Hilbert space follows a unitary operator determined by the Hamiltonian of the system. Finally, when a measurement is performed on a system, an eigenvalue of the measured variable (operator) is found, and the state is projected on the corresponding eigenspace. Although these concepts seem at odds with physical laws at our scale, the quantum rules do lead to the classical behavior for a system coupled to a sufficiently complex environment. More precisely, the theory of decoherence in quantum mechanics predicts that the entanglement between the system and its environment suppresses coherence between system states (interferences are no longer possible), and yields probabilities for the states that can result from the evolution. Classical physics does not derive from quantum mechanics in the sense that the state emerging from the evolution of the system coupled to its environment is predicted only statistically. As a result, quantum physics has been mainly considered as relevant for the description of the microscopic world, although no distinction exists in principle between various kinds of degrees of freedom: their underlying complexity does not come into play within the standard framework of quantum mechanics.

This blindness explains the fifty years delay between the establishment of quantum mechanics, and the first proposals of quantum machines in the nineteen-eighties. On the experimental side, the investigation of quantum effects in electronic circuits carried out during the last thirty years paved the way to this conceptual revolution. The question of the quantumness of a collective variable involving a large number of microscopic particles, such as the current in a superconducting circuit, was raised. The quantitative observation of quantum effects such as macroscopic quantum tunneling [8] contributed to establishing the confidence that quantum mechanics can be brought in the realm of macroscopic objects.

Before embarking on the description of qubits, it is worth noticing that quantum machines offer a new direction to probe quantum mechanics. Recently, the emphasis has been put on the entanglement degree rather than on the mere size of a quantum system. Probing entanglement between states of macroscopic circuits, or reaching quantum states with a high degree of entanglement are now major issues in quantum physics. This is the new border, whose exploration started by the demonstration of the violation of Bell's inequalities for entangled pairs of photons [9]. This research direction, confined for a long time in Byzantine discussions about the EPR and Schrödinger cat paradoxes, is now accesible to experimental tests [10].

*First proposals of quantum machines* Commonly accepted quantum machines such as the laser only involve quantum mechanics at the microscopic level, atom-field interactions in this case. A true quantum machine is a system in which machine-state variables are ruled by quantum mechanics. One might think that quantum machines more complex than molecules could not exist because the interactions between any complex system and the numerous degrees of freedom of its environment tend to drive it into the classical regime. Proposing machines that could benefit from the quantum rules was thus a bold idea. Such propositions appeared in the domain of processors after Deutsch and Josza showed that the concept of algorithmic complexity is hardware dependent. More precisely, it was proved that a simple set of unitary operations on an ensemble of coupled two level systems, called qubits, is sufficient to perform some specific computing tasks in a smaller number of algorithmic steps than with a classical processor [1].

Although the first problem solved "more efficiently" by a quantum algorithm was not of great interest, it initiated great discoveries. Important results were obtained [1], culminating with the factorisation algorithm discovered by Shor in 1994, and with the quantum error correction codes [1] developed by Shor, Steane, Gottesman and others around 1996. These breakthroughs should not hide the fact that the number of quantum algorithms is rather small. But since many problems in the same complexity class can be equivalent, solving one of them can provide a solution to a whole class of problems. Pessimists see in this lack of algorithms a major objection to quantum computing. Optimists point out that simply to simulate quantum systems, it is already worthwhile to develop quantum processors, since this task is notoriously difficult for usual computers. A more balanced opinion might be that more theoretical breakthroughs are still needed before quantum algorithms are really worth the effort of making quantum processors. How large does a quantum processor need to be to perform a useful computation? It is considered that a few tens of robust qubits would already be sufficient for performing interesting computations. Notice that the size of the Hilbert space of such a processor is already extremely large.

## 1.2. *Quantum processors based on qubits*

A sketch of a quantum processor based on quantum bits is shown in Fig. 1. It consists of an array of these qubits, which are two level systems. Each qubit is controlled independently, so that any unitary operation can be applied to it. Qubits are coupled in a controlled way so that all the two qubit gate operations required by algorithms can be performed. As in Boolean logic, a small set of gates is sufficient to form a universal set of operations, and hence to operate a quantum processor. A two-qubit gate is universal when, combined with a subset of single qubit gates, it allows implementation of any unitary evolution [1]. For instance, the control-not gate (C-NOT), which applies a not operation on qubit 2 when qubit1 is in state 1, is universal.

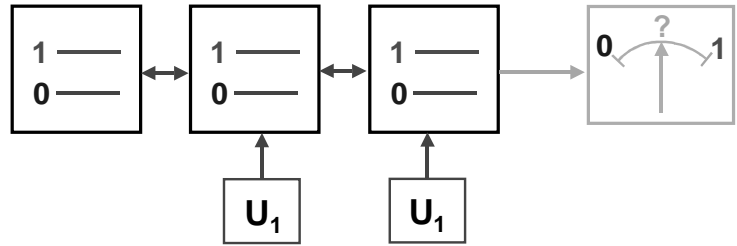

Fig. 1. Sketch of a quantum processor based on qubits. Each qubit is here a robust qubit, with its error correction circuitry. The detailed architecture of a quantum processor strongly depends on the set of gates that can be implemented. The single two qubit gates, combined with single qubit gates, should form a universal set of gates, able to process any quantum algorithm.

*Criteria required for qubits*   Not all two level systems are suitable for implementing qubits. A series of points, summarised by DiVicenzo, need to be addressed (see chapter 7 in [1]):

1) The level spectrum should be sufficiently anharmonic to provide a good two level system.

2) An operation corresponding to a 'reset' is needed.

3) The quantum coherence time must be sufficient for the implementation of quantum error correction codes.

4) The qubits must be of a scalable design with a universal set of gates.

5) A high fidelity readout method is needed.

These points deserve further comments:

The requirement on the coherence is measured by the number of gate operations that can be performed with an error small enough so that error correcting codes can be used. This requirement is extremely demanding: less than one error in $10^4$ gate operations. Qubits rather better protected from decoherence than those available today will be needed for this purpose.

If a readout step is performed while running the algorithm, a perfect readout system should provide answers with the correct probabilities, and project the register on the state corresponding to the outcome read.The state can then be stored for other purpose. This is the definition of a quantum non demolition (QND) measurement. Such a QND readout would be useful to measure quantum correlations in coupled qubit circuits and to probe whether or not Bell's inequalities are violated as predicted by quantum mechanics like in the microscopic world [9]. However, non QND readout systems could provide answers with the correct probabilities, but fail to achieve the projection afterwards. Note that QND readout is not essential for quantum algorithms; although the factorization algorithm is often presented with intermediate projection step, it is not necessary.

### 1.3. Atom and ion versus solid state qubits

On the experimental side, implementing quantum processors is a formidable task, and no realistic scalable design presently exists. The activity has been focused on the operation of simple systems, with at most a few qubits. Two main roads have been followed. First, microscopic quantum systems like atoms [10] and ions [11] have been considered. Their main advantage is their excellent quantumness, but their scalability is questionable. The most advanced qubit implementation is based on ions in linear traps, coupled to their longitudinal motion [11] and addressed optically. Although the trend is to develop atom-chips, these implementations based on microscopic quantum objects still lack the flexibility of microfabricated electronic circuits, which constitute the second main road investigated. Here, quantumness is limited by the complexity of the circuits that always involve a macroscopic number of atoms and electrons. We describe in the following this quest for quantumness in electronic solid state circuits.

### 1.4. Electronic qubits

Two main strategies based on quantum states of either single particles or of the whole circuit, have been followed for making solid-state qubits.

In the first strategy, the quantum states are nuclear spin states, single electron spin states, or single electron orbital states. The advantage of using microscopic states is that their quantumness has already been probed and can be good at low temperature. The main drawback is that qubit operations are difficult to perform since single particles are not easily controlled and read out.

The second strategy has been developed in superconducting circuits based on Josephson junctions, which form a kind of artificial atoms. Their Hamiltonian can be tailored almost at will, and a direct electrical readout can be incorporated in the circuit. On the other hand, the quantumness of these artificial atoms does not yet compare to that of natural atoms or of spins.

## 2. qubits in semiconductor structures

Microscopic quantum states suitable for making qubits can be found in semiconductor nanostructures, but more exotic possibilities such as Andreev states at a superconducting quantum point contact [12] have also been proposed. Single particle quantum states with the best quantumness have been selected, and a few representative approaches are described below. Two families can be distinguished: the first one being based on quantum states of nuclear spins, or of localised electrons, while the second one is based on propagating electronic states (flying qubits).

### 2.1. Kane's proposal: nuclear spins of P impurities in silicon

The qubits proposed by Kane are the S=1/2 nuclear spins of $P^{31}$ impurities in silicon [13]. Their quantumness is excellent, and rivals that of atoms in vacuum. In the ref. [13], the author has proposed a scheme to control, couple and readout such spins. A huge effort has been started in Australia in order to implement this proposal sketched in Fig. 2. The qubits are controlled through the hyperfine interaction between the nucleus of the $P^{31}$ impurity and the bound electron around it. The effective Hamiltonian of two neighboring nucleus bound electron spins:

$$H = A_1 \sigma^{1n} \sigma^{1e} + A_2 \sigma^{1n} \sigma^{1e} + J \sigma^{1e} \sigma^{2e},$$

where the subscripts $n$ and $e$ refer to nuclei and bound electrons respectively. The transition frequency of each qubit is determined by the magnetic field applied to it, and by its hyperfine coupling $A$ controlled by the gate voltage applied to the A gate electrode, which displaces the wavefunction of the bound electron. Single qubit gates would be performed by using resonant pulses, like in NMR, while two qubit gates would be performed using the J gates, which tune the exchange interaction between neighboring bound electrons. The readout would be performed by transfering the information on the qubit state to the charge of a quantum dot, which would then be read using an rf-SET. Although the feasibility of Kane's proposal has not yet been demonstrated, it has already yielded significant progress in high accuracy positioning of a single impurity atom inside a nanostructure.

### 2.2. Electron spins in quantum dots

Using electron spins for the qubits is attractive because the spin is weakly coupled to the other degrees of freedom of the circuit, and because the spin state can be transferred to a charge state for the purpose of readout (see [14] and refs. therein).

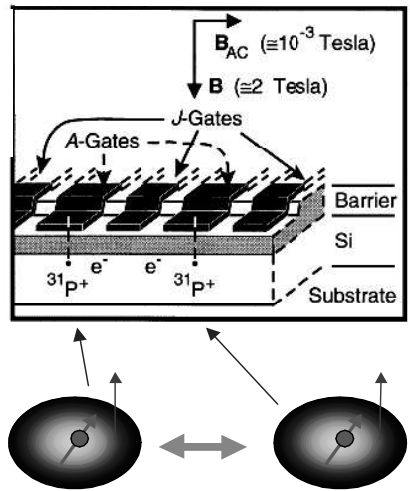

Fig. 2. Kane's proposal: nuclear spins of phosphorus impurities form the qubits.The control is provided by the hyperfine interaction with a bound electron around each impurity. Each qubit level scheme is controlled by applying a voltage to an A gate (labelled A) electrode that displaces slightly the wavefunction of the bound electron, and thus modifies the hyperfine interaction. Single qubit gates are performed by applying an ac field on resonance, like in Nuclear Magnetic Resonance. The two qubit gates are performed using the J gates (labelled J), which control the exchange interaction between neighboring bound electrons. The exchange interaction mediates an effective interaction between the qubits. The readout is performed by transfering the information on the qubit state to the charge of a quantum dot (not shown), which is then read using an rf-Single Electron Transistor *(picture taken from [13].)*

Single qubit operations can be performed by applying resonant magnetic fields (ESR), and two qubit gates can be obtained by controlling the exchange interaction between two neighboring electrons in a nanostructure. The device shown in Fig. 3 is a double dot in which the exchange interaction between the single electrons in the dots is controlled by the central gate voltage. The readout is performed by monitoring the charge of the dot with a quantum point contact transistor close to it. The measurement proceeds as follows: first, the dot gate voltage is changed so that an up spin electron stays in the dot, while a down spin electron leaves it. In that case, another up spin electron from the reservoir can enter the dot. The detection of changes in the dot charge thus provides a measurement of the qubit state. Note that such a measurement can have a good fidelity as required, but is not QND because the quantum state is destroyed afterwards.

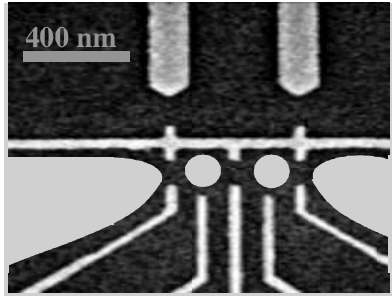

Fig. 3. Scanning Electron Micrograph of a double dot implementing two qubits . The qubits are based on the spin of a single electron in the ground state of each dot (disks). The two qubits are coupled through the exchange energy between electrons, which is controlled by the central gate. Single qubit gates are obtained by applying local resonant ac magnetic fields. Readout is performed by monitoring each dot charge with a point contact transistor, after a sudden change of the dot gate voltage. An electron with the up spin state stays in the dot, whereas a down spin exits, and is replaced by an up spin electron. A change in the dot charge thus signals a down spin. *(Courtesy of Lieven Vandersypen, T.U. Delft).*

### 2.3. Charge states in quantum dots

The occupation of a quantum dot by a single electron is not expected to provide an excellent qubit because the electron strongly interacts with the electric field. Coherent oscillations in a semiconductor qubit circuit [15] were nevertheless observed by measuring the transport current in a double dot charge qubit repetitively excited by dc pulses, as shown in Fig. 4.

### 2.4. Flying qubits

Propagating electron states provide an interesting alternative to localised states. Propagating states in wires with a small number of conduction channels have been considered, but edge states in Quantum hall Effect structures offer a better solution [4] . Due to the absence of back-scattering, the phase coherence time at low temperature is indeed long: electrons propagate coherently over distances longer than $100 \ \mu m$. Qubit states are encoded using electrons propagating in opposite directions, along the opposite sides of the wires. The qubit initialisation can be performed by injecting a single electron in an edge state. As shown in Fig. 5, single qubit gates can be obtained with a quantum point contact that transmits or reflects incoming electrons, and two qubit gates can be obtained by coupling edge states over a short length. The readout can be performed by detecting the passage of the electrons along the wire, using a corrugated edge in order to increase the readout time. This system is not easily scalable because of its topology, but is well suited for entangling pairs of electrons and measuring their correlations.

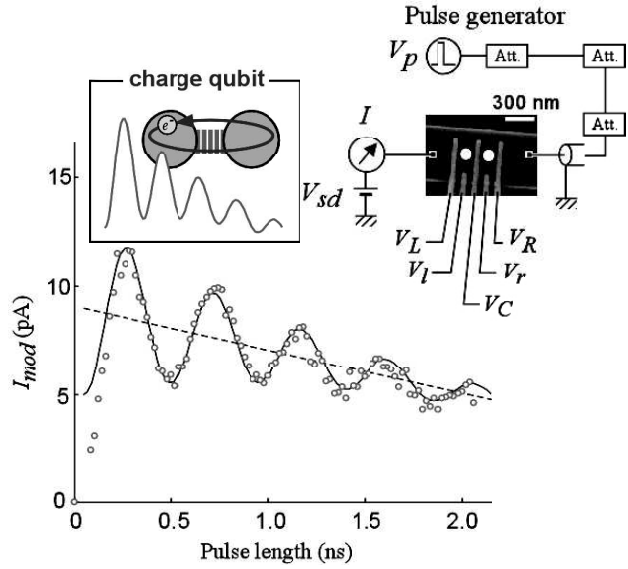

Fig. 4. Coherent oscillations of a single electron inside a double dot structure, as a function of the duration of a dc pulse applied to the transport voltage. These oscillations are revealed by the average current when the pulse is repeated at a large rate *(picture taken from Hayashi et al. [15] )*

## 3. Superconducting qubit circuits

The interest of using the quantum states of a whole circuit for implementing qubits is to benefit from the wide range of Hamiltonians that can be obtained when inductors and capacitors are combined with Josephson junctions. These junctions are necessary because a circuit built solely from inductors and capacitors constitutes a set of harmonic modes. A Josephson junction [16] has a Hamiltonian which is not quadratic in the electromagnetic variables, and hence allows to obtain an anharmonic energy spectrum suitable for a qubit. Josephson qubits can be considered as artificial macroscopic atoms, whose properties can be tailored. The internal and coupling Hamiltonians can be controlled by applying electric or magnetic fields, and bias currents. The qubit readout can also be performed electrically.

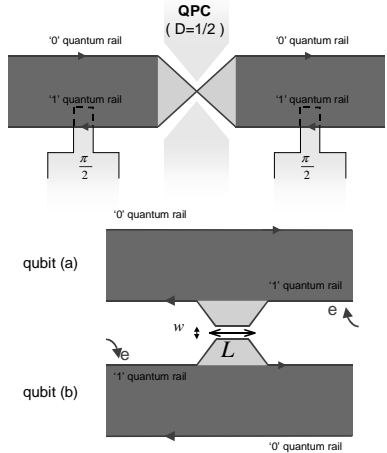

Fig. 5. Single qubit gate (top) and two qubit gate (bottom) for flying qubits based on edge states in QHE nanostructures *(Courtesy of C. Glattli)*.

### 3.1. *Josephson qubits*

A direct derivation of the Hamiltonian can often be performed for simple circuits. There are however systematic rules to derive the Hamiltonian of a Josephson circuit [17, 18], and different forms are possible depending on the choice of variables. When branch variables are chosen, the contribution to the Hamiltonian of a Josephson junction in a given branch is:

$$h(\theta) = -E_J \cos(\theta),$$

where $E_J = I_0 \phi_0$ is the Josephson energy, with $I_0$ the critical current of the junction, and $\theta$ the superconducting phase difference between the two nodes connected by the branch. The phase $\theta$ is the conjugate of the number $N$ of Cooper pairs passed across the junction. In each quantum state of the circuit, each junction is characterised by the fluctuations of $\theta$ and of $N$. Often, the circuit junctions are either in the phase or number regimes, characterised by small and large fluctuations of the phase, respectively. Qubit circuits can be classified according to the regime to which they belong.

### 3.1.1. Hamiltonian of Josephson qubit circuits

In the case of a single junction, the electromagnetic Hamiltonian of the circuit in which the junction is embedded adds to the junction Hamiltonian. The phase biased junction is in the phase regime, whereas the charge biased junction, a circuit called the Single Cooper Pair Box, can be in a charge regime, phase regime, or intermediate charge-phase regime, depending on the circuit parameters. The Cooper-pair box in the charge regime was the first Josephson qubit in which coherent behavior was demonstrated [19].

In practice, all Josephson qubits are multi junction circuits in order to tailor the Hamiltonian, to perform the readout, and to achieve the longest possible coherence times. The main types of superconducting qubit circuits can be classified along a phase to charge axis, as shown in Fig. 6. The phase qubit [20] developed at NIST (Boulder) consists of a Josephson junction in a flux biased loop, with two potential wells. The qubit states are two quantized levels in the first potential well, and the readout is performed by resonantly inducing the transfer to the second well, using a monitoring SQUID to detect it. The flux qubit [21, 22] developed at T.U. Delft consists of three junctions in a loop, placed in the phase regime. Its Hamiltonian is controlled by the flux threading the loop. The flux qubit can be coupled in different ways to a readout SQUID. The quantronium circuit [7, 23–25], developed at CEA-Saclay is derived from the Cooper pair box, but is operated in the intermediate charge-phase regime. A detailed description of all Josephson qubits, with extensive references to other works, is given in [5–7].

### 3.1.2. The single Cooper pair box

The single Cooper pair box [7] consists of a single junction connected to a voltage source across a small gate capacitor, as shown in Fig. 7. Its Hamiltonian is the sum of the Josephson Hamiltonian and of an electrostatic term:

$$\widehat{H}(N_g) = E_C(\widehat{N} - N_g)^2 - E_J \cos\widehat{\theta}\,, \tag{3.1}$$

where $E_C = (2e)^2/2C_\Sigma$ is the charging energy, and $N_g = C_g V_g/(2e)$ the reduced gate charge with $V_g$ the gate voltage. The operators $\widehat{N}$ and $\widehat{\theta}$ obey the commutation relation $\left[\widehat{\theta}, \widehat{N}\right] = i$. The eigenstates and eigenenergies can be analytically determined, or calculated numerically using a restriction of the Hamiltonian in a subspace spanned by a small set of $|N\rangle$ states. They are $2e$ periodic with the gate charge. The two lowest energy levels provide a quantum bit because the eigenenergy spectrum is anharmonic for a wide range of parameters. When $E_J \ll E_C$, the qubit states are two successive $|\mathbf{N}\rangle$ states away from $N_g \equiv 1/2 \bmod[1]$, and symmetric and antisymmetric combinations of successive $|N\rangle$ states in the vicinity of $N_g \equiv 1/2 \bmod[1]$.

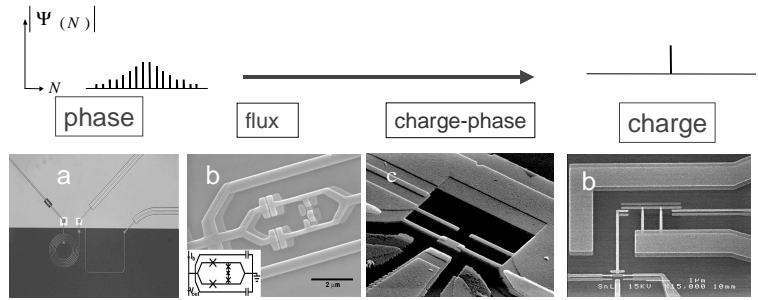

Fig. 6. Josephson qubits can be classified along an axis ranging from the phase regime to the charge regime: the current biased large junction(a), the flux qubit (b), the quantronium charge-phase qubit (c), the Cooper pair box with small Josephson energy (d). In the phase regime, the number of Cooper pairs transferred across each junction has large fluctuations, whereas these fluctuations are small in the charge regime. *(Courtesy of NIST, T.U. Delft, CEA-Saclay, and Chalmers).*

### 3.1.3. Survey of Cooper pair box experiments

The most direct way to probe the Cooper pair box is to measure the island charge. Following this idea, the island charge was measured in its ground state in 1996 [26] by capacitively coupling the box island to an electrometer based on a Single Electron Transistor (SET) [27]. This readout method could not be used however for time resolved experiments because its measuring time was too long. The first Josephson qubit experiment was performed in 1999 at NEC [19], by monitoring the current through an extra junction connected on one side to the box island and on the other side to a voltage source. When the box gate charge is suddenly (i.e. non adiabatically) moved from $N_g \approx 0$ to $N_g = 1/2$, the initial ground state $|0\rangle$ state is no longer an eigenstate, and coherent oscillations between states take place between $|0\rangle$ and $|1\rangle$ at the qubit transition frequency. When $N_g$ is suddenly moved back to its initial value $\approx 0$, the probability for the qubit to be in the excited state $|1\rangle$ is conserved. The readout takes advantage of the available energy in the upper state to transfer a Cooper pair across the readout junction. When the experiment is repeated, the average current through the readout junction provides a measurement of the qubit state at the end of the gate charge pulse. This method of readout provides a continuous average measurement of the box. It proved extremely well suited to many experiments. However, it cannot provide a single shot readout of the qubit. The evolution of qubit design was then driven by the aim of achieving a better quantumness and a more efficient readout. Better quantumness means a longer coherence time, with a controlled influence of the environment to avoid decoherence. More efficient readout means single shot readout, with a fidelity as high as possible, and ideally quantum non demolition (QND). The quantronium operated in 2001 at Saclay was the first qubit circuit combining a single shot readout with a long coherence time [7, 23–25]. In 2003, the charge readout of a Cooper pair box was achieved at Chalmers [29] using an rf-SET [30], which is a SET probed at high frequency. A sample and hold charge readout was operated in 2004 at NEC [31], with a fidelity approaching 90%. In 2004, a Cooper pair box embedded in a resonant microwave cavity was operated at Yale [32] using the modification of the cavity transmission by the Cooper pair box, similar to the effect of a single atom in cavity-QED experiments [10].

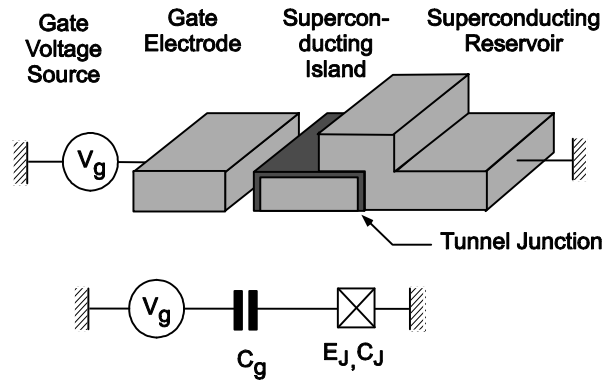

Fig. 7. Schematic representation and electrical circuit of a Single Cooper pair box: A small super-conducting island is connected to a voltage source across a capacitor on one side, and a Josephson junction on the other side. In the schematic circuit, the cross in a box represents a small Josephson junction.

## 3.2. *How to maintain quantum coherence?*

When the readout circuit measures the qubit, its backaction results in full qubit decoherence during the time needed to get the outcome, and even faster if the readout efficiency is below the quantum limit. The readout should thus be switched off when the qubit is operated, and switched on just at readout time. Furthermore, even when the readout is off, the qubit is subject to decoherence, partly due to the connection of the qubit to the readout circuitry. How could one possibly limit the influence of the environment and of the readout even when it is off, and to switch on the readout when needed?

Before explaining a possible strategy to circumvent this major problem, we expose the basic concepts underlying decoherence in qubit circuits. The interaction between a qubit and the degrees of freedom of its environment entangles both parties. This entanglement takes a simple form in the weak coupling regime, which is usually the case in qubit circuits [28]. The coupling arises from the fact that the control parameters of the qubit Hamiltonian ( such as $N_g$ for the Cooper pair box), are in fact fluctuating variables of the qubit environment.

### 3.2.1. *Qubit-environment coupling Hamiltonian*

We call $\lambda$ the set of control variables entering the Hamiltonian of a qubit. At a given working point $\lambda_0$, the qubit space is analogous to a fictitious spin $1/2$ with $\sigma_z$ eigenstates $|0\rangle$ and $|1\rangle$. Using the Pauli matrix representation of spin operators, the expansion of the Hamiltonian around $\lambda_0$ yields the coupling Hamiltonian:

$$\widehat{H}_X = -1/2 \left( \overrightarrow{D_\lambda} \cdot \overrightarrow{\sigma} \right) \left( \widehat{\lambda} - \lambda_0 \right), \tag{3.2}$$

where $\overrightarrow{D_\lambda} \cdot \overrightarrow{\sigma}$ is the restriction of $-2\widehat{\partial H / \partial \lambda}$ to the $\{|0\rangle, |1\rangle\}$ space. This coupling Hamiltonian determines the qubit evolution when a control parameter is varied at the qubit transition frequency, and the coupling to decoherence noise sources.

In the weak coupling regime, the fluctuations of the qubit environment are characterised by the spectral density:

$$S_{\lambda_0}(\omega) = \frac{1}{2\pi} \int_{-\infty}^{+\infty} d\tau \left\langle \left( \widehat{\lambda}(t) - \lambda_0 \right) \left( \widehat{\lambda}(t+\tau) - \lambda_0 \right) \right\rangle \exp(-\mathbf{i}\omega\tau) \tag{3.3}$$

This spectral density is defined for positive and negative $\omega'$s, proportional to the number of environmental modes that can absorb and emit a quantum $\hbar\omega$, respectively. In the case of the Cooper pair box, the fluctuations of the gate charge $N_g$ arise from the impedance of the biasing circuitry and from microscopic charge fluctuators in the vicinity of the box island [7, 25].

### 3.2.2. *Relaxation*

The decay of the longitudinal part of the density matrix in the eigenstate basis $\{|0\rangle, |1\rangle\}$ involves $|1\rangle \rightarrow |0\rangle$ qubit transitions, with the energy transferred to the

environment. Such an event resets the qubit in its ground state. The decay is exponential, with a rate:

$$\Gamma_1 = \frac{\pi}{2} \left( \frac{D_{\lambda,\perp}}{\hbar} \right)^2 S_{\lambda_0}(\omega_{01}) \ . \tag{3.4}$$

The symbol $\perp$ indicates that only transverse fluctuations at positive frequency $\omega_{01}$ induce downward transitions. Upward transitions, which involve $S_{\lambda_0}(-\omega_{01})$, occur at a negligible rate for experiments performed at temperatures $k_B T \ll \hbar\omega_{01}$, provided the environment is at thermal equilibrium. The relaxation time is thus $T_1 = 1/\Gamma_1$.

### 3.2.3. Decoherence: relaxation + dephasing

When a coherent superposition $a\,|0\rangle + b\,|1\rangle$ is prepared, the amplitudes $a$ and $b$ evolve in time, and the non diagonal part of the density matrix oscillates at the qubit frequency $\omega_{01}$. The precise definition of decoherence is the decay of this part of the density matrix. There are two distinct contributions to this decay. Relaxation contributes to decoherence by an exponential damping factor with a rate $\Gamma_1/2$. Another process, called dephasing, often dominates. When the qubit frequency $\Omega_{01}$ fluctuates, an extra phase factor $\exp[i\Delta\varphi(t)]$ with $\Delta\varphi(t) = \frac{D_{\lambda,z}}{\hbar} \int_0^t \left( \widehat{\lambda}(t') - \lambda_0 \right) dt'$ builds up between both amplitudes, the coupling coefficient $D_{\lambda,z}$ being:

$$D_{\lambda,z} = \langle 0|\, \widehat{\partial H/\partial \lambda}\,|0\rangle - \langle 1|\, \widehat{\partial H/\partial \lambda}\,|1\rangle = \hbar\partial\omega_{01}/\partial\lambda \ .$$

Dephasing involves longitudinal fluctuations, and contributes to decoherence by the factor:

$$f_X(t) = \langle \exp[i\Delta\varphi(t)]\rangle \ . \tag{3.5}$$

Note that the decay of this dephasing factor $f_X(t)$ is not necessarily exponential. When $D_{\lambda,z} \neq 0$, and assuming a gaussian process for $\left( \widehat{\lambda}(t') - \lambda_0 \right)$, one finds using a semi-classical approach:

$$f_X(t) = \exp\left[ -\frac{t^2}{2} \left( \frac{D_{\lambda,z}}{\hbar} \right)^2 \int_{-\infty}^{+\infty} d\omega\, S_{\lambda_0}(\omega)\mathrm{sinc}^2(\frac{\omega t}{2}) \right] \ , \tag{3.6}$$

A full quantum treatment of the coupling to a bath of harmonic oscillators justifies using the quantum spectral density in the above expression [28]. When the spectral density $S_{\lambda_0}(\omega)$ is regular at $\omega = 0$, and flat at low frequencies, $f_X(t)$ decays exponentially at long times, with a rate $\Gamma_\varphi \approx \pi \left( \frac{D_{\lambda,z}}{\hbar} \right)^2 S_{\lambda_0}(\omega \approx 0)$.

When the spectral density diverges at $\omega = 0$, like for the ubiquitous $1/f$ noise, a careful evaluation has to be performed [33, 34].

### 3.2.4. The optimal working point strategy

The above considerations on decoherence yield the following requirements for the working point of a qubit:

-In order to minimize the relaxation, the coefficients $D_{\lambda,\perp}$ should be small, and ideally $D_{\lambda,\perp} = 0$.

-In order to minimize dephasing, the coefficients $D_{\lambda,z} \propto \partial\Omega_{01}/\partial\lambda$ should be small. The optimal case is when the transition frequency is stationary with respect to all control parameters: $D_{\lambda,z} = 0$. At such optimal points, the qubit is decoupled from its environment and from the readout circuitry in particular. This means that the two qubit states cannot be discriminated at an optimal point. One must therefore depart in some way from the optimal point in order to perform the readout. The first application of the optimal working point strategy was applied to the Cooper pair box, with the quantronium circuit [7, 23, 24].

## 4. The quantronium circuit

The quantronium circuit is derived from a Cooper pair box. Its Josephson junction is split into two junctions with respective Josephson energies $E_J(1 \pm d)/2$, with $d \in [0, 1]$ a small asymmetry coefficient (see Fig. 8). The reason for splitting the junction into two halves is to form a loop that can be biased by a magnetic flux $\Phi$. The split box, which we first explain, has two degrees of freedom, which can be chosen as the island phase $\widehat{\theta}$ and the phase difference $\widehat{\delta}$ across the two box junctions. In this circuit, the phase $\widehat{\delta}$ is a mere parameter $\delta = \Phi/\phi_0$.

The Hamiltonian of the split box, which depends on the two control parameters $N_g$ and $\delta$, is:

$$\widehat{H} = E_C(\widehat{\mathbf{N}} - N_g)^2 - E_J \cos(\frac{\widehat{\delta}}{2})\cos(\widehat{\theta}) + dE_J \sin(\frac{\widehat{\delta}}{2})\sin(\widehat{\theta}) \,. \qquad (4.1)$$

The two lowest energies of this Hamiltonian are shown in Fig. 9 as a function of the control parameters. The interest of the loop is to provide a new variable to probe the qubit: the loop current. The loop current is defined by the operator:

$$\widehat{I}(N_g, \delta) = (-2e)\left(-\frac{1}{\hbar}\frac{\partial\widehat{H}}{\partial\delta}\right) \,.$$

The average loop current $\langle i_k \rangle$ in state $|k\rangle$ obeys a generalized Josephson relation: $\langle i_k(N_g, \delta) \rangle = \left\langle k \left| \widehat{I} \right| k \right\rangle = \frac{1}{\varphi_0}\partial E_k(N_g, \delta)/\partial\delta$ . The difference between the loop

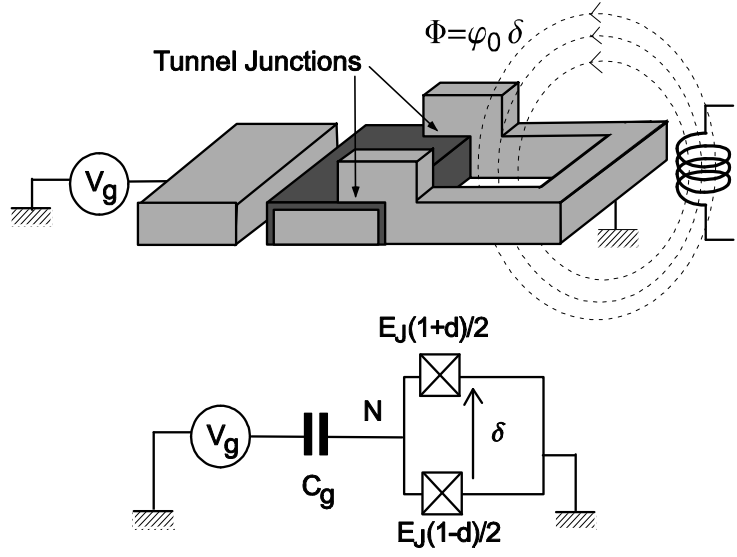

Fig. 8. *Schematic representation of a split Cooper pair box showing its island, its two Josephson junctions connected to form a grounded superconducting loop, its gate circuit, and its magnetic flux bias. Bottom: Corresponding electrical drawing.*

currents of the two qubit states is $\Delta i_{10} = \langle i_1 \rangle - \langle i_0 \rangle = 2e\partial\omega_{10}/\partial\delta$. As expected, the difference $\Delta i_{10}$ vanishes at an optimal point.

The variations of the qubit transition frequency with the control parameters are shown in Fig.10. Different optimal points where all derivatives $\partial\Omega_{01}/\partial\lambda_i$ vanish are present. The charge difference $\Delta N_{10} = \langle N_1 \rangle - \langle N_0 \rangle$ also vanishes at these points. The optimal point $\{N_g = 1/2, \delta = 0\}$ was first used.

## 4.1. Relaxation and dephasing in the quantronium

The split box is unavoidably coupled to noise sources affecting the gate charge $N_g$ and the phase $\delta$ [7, 25]. The coupling to these noise sources $D_{\lambda,\perp}$ and $D_{\lambda,z}$ for relaxation and dephasing are obtained from the definition 3.2.

The coupling vector $D_{\lambda,\perp}$ for relaxation is:

$D_{\lambda,\perp} = \left\{ 4E_C \left| \left\langle 0 \left| \widehat{N} \right| 1 \right\rangle \right|, 2\varphi_0 \left| \left\langle 0 \left| \widehat{I} \right| 1 \right\rangle \right| \right\}$.

Relaxation can thus proceed through the charge and phase ports, but one finds that the phase port does not contribute to relaxation at $N_g = 1/2$ when the asymmetry factor $d$ vanishes. Precise balancing of the box junctions is thus important in the quantronium.

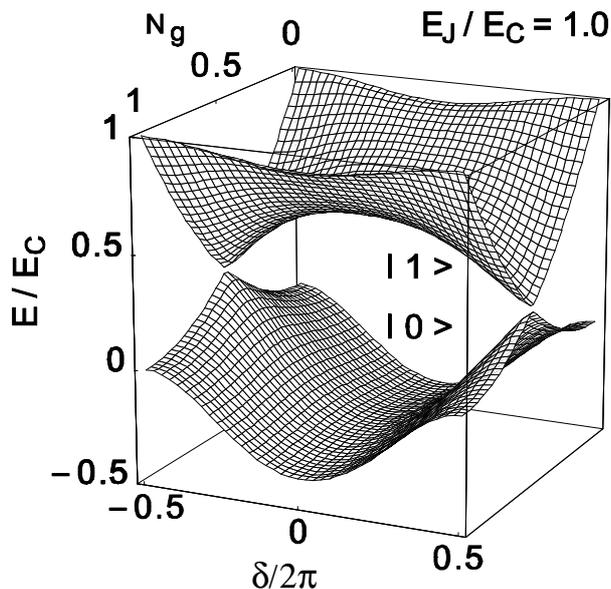

Fig. 9. Two lowest energy levels of a split Cooper pair box having an $E_J/E_C$ ratio equal to 1, as a function of the two external parameters $N_g$ and $\delta$. Energies are normalized by the Cooper pair Coulomb energy. The asymetry coefficient used here is $d = 2\%$. The principal effect of $d$ is to control the gap at $N_g = 1/2$, $\delta = \pi$.

The coupling vector for dephasing is directly related to the derivatives of the transition frequency:

$D_{\lambda,z} = \hbar \; (\partial\omega_{01}/\partial N_g, \; \partial\omega_{01}/\partial\delta)$.

The charge noise arises from the noise in the gate bias circuit and from the background charge noise due to microscopic fluctuators in the vicinity of the box tunnel junctions. This background charge noise has a $1/f$ spectral density at low frequency, with a rather universal amplitude. The phase noise also has a $1/f$ spectral density, but its origin in Josephson junction circuits is not well understood and is not universal.

### 4.2. Readout

The full quantronium circuit, shown in the top of Fig. 11, consists of a split-box with an extra larger junction inserted in the loop for the purpose of readout. The Hamiltonian of the whole circuit is the sum of the split-box Hamiltonian4.1 and of the Hamiltonian of a current-biased Josephson junction [7, 25]. The phase difference $\widehat{\delta}$ in the split-box Hamiltonian is related to the phase difference across the readout junction by the relation $\widehat{\delta} = \widehat{\gamma} + \Phi/\phi_0$. The phase $\widehat{\delta}$ is still an almost classical variable, except at readout time, when the qubit gets entangled with the readout junction. This readout junction can be used in different ways in order to discriminate the qubit states, as we now show.

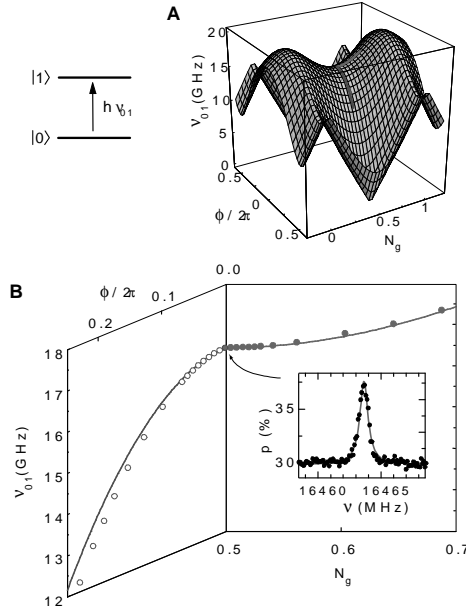

Fig. 10. A: Calculated transition frequency as a function of the control parameters $N_g$ and $\delta$ for the parameters $E_J = 0.86 \ k_B K$, $E_C = 0.68 \ k_B K$. Three optimal points where the frequency is stationary, are visible. The optimal point used in the experiments is the saddle point ($N_g = 1/2, \delta = 0$) . B: cuts along the planes $N_g = 1/2$ and $\delta = 0$. Symbols: position of the resonance of the switching probability in CW excitation; lines: predictions. The lineshape at the optimal point is plotted in inset.( *Taken from Vion et al. [24]*).

### 4.2.1. Switching readout

The simplest readout method consists in using the readout junction to perform a measurement of the loop current after adiabatically moving away from the optimal point. For this purpose, a trapezoidal readout pulse with a peak value slightly below the readout junction critical current is applied to the circuit. Since this bias current adds to the loop current in the readout junction, the switching of the readout junction to a finite voltage state can be induced with a large probability for state $|1\rangle$ and with a small probability for state $|0\rangle$). This switching method is in principle a single shot readout. It has been applied to the quantronium [**?**] and to the flux qubit [22], with switching probability difference up to $40\%$ and $70\%$, respectively. The lack of fidelity is attributed to spurious relaxation during the readout bias current pulse.This switching method also destroys the qubit after measurement: this is not a QND readout.

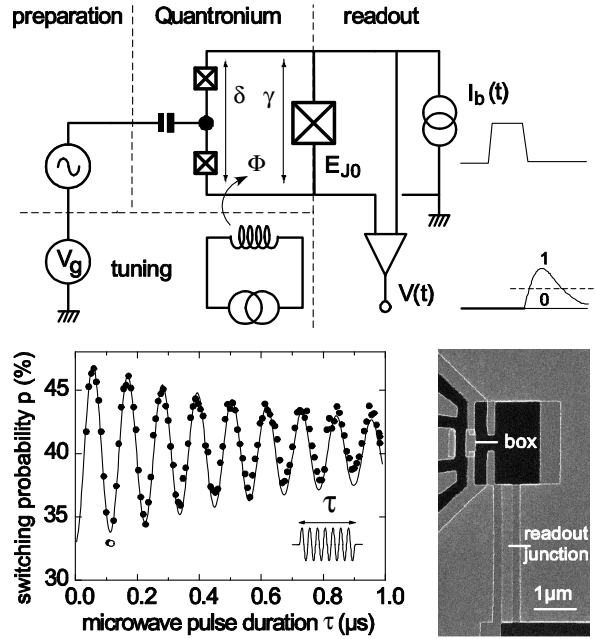

Fig. 11. Top: Schematic circuit of the quantronium qubit circuit. The quantronium consists of a readout junction inserted in the loop of a split-junction Cooper pair box. When a trapezoidal current pulse is applied, the readout junction switches to the voltage state with a larger probability for state $|1\rangle$ than for state $|0\rangle$. Bottom right: Scanning electron picture of a quantronium sample made using double angle shadow-mask evaporation of aluminum. Bottom left: Rabi oscillations of the switching probability as a function of the duration of a resonant microwave pulse.

### 4.2.2. AC methods for QND readout

Recently, microwave methods measuring the phase of a microwave signal reflected or transmitted by the circuit have been used with various superconducting qubits in order to attempt a non destructive readout. A QND readout should also lead to a better readout fidelity. Although correlated measurements on coupled qubits and quantum algorithms do not require QND readout, achieving this goal seems essential for probing quantum mechanics in macroscopic objects. In general, with these rf methods, the working point stays, on average, at the optimal point, and undergoes small amplitude oscillations at a frequency different than the qubit frequency. The case of the Cooper pair box embedded in a resonant microwave cavity is an exception because the cavity frequency is comparable with the qubit frequency [32]. Avoiding moving far away from the optimal point might thus reduce the spurious relaxation observed with the switching method, and thus improve the readout fidelity. These rf methods, proposed for the flux qubit [35], the quantronium [36,37], and the Cooper pair box [38], give access to the second derivative of the energy of each qubit state with trespect to the control parameter that is driven. In the quantronium, this parameter is the phase $\hat{\delta}$. The qubit slightly modifies the inductance of the whole circuit [37], with opposite changes for the two qubit states. The readout of the inductance change is obtained by measuring the reflected signal at a frequency slightly below the plasma frequency of the readout junction. The discrimination between the two qubit states is furthermore greatly helped there by the non-linear resonance of the junction and the consequent dynamical transition from an in phase oscillation regime, to an out of phase oscillation regime when the drive amplitude is increased [36].

## 5. Coherent control of the qubit

Coherent control of a qubit is performed by driving the control parameters of the Hamiltonian. This evolution of the qubit state can be either adiabatic, or non adiabatic. A slow change of the control parameters yields an adiabatic evolution of the qubit that can be useful for some particular manipulations. Note however that the adiabatic evolution of the ground state of a quantum system can be used to perform certain quantum computing tasks [39]. Two types of non-adiabatic evolutions have been performed, with dc-pulses and with resonant ac-pulses.

### 5.1. Ultrafast 'DC' pulses versus resonant microwave pulses

In the dc-pulse method [19], a sudden change of the Hamiltonian is performed. The qubit state does not in principle evolve during the change, but evolves afterwards with the new Hamiltonian. After a controlled duration, a sudden return to the initial working point is performed in order to measure the qubit state. In this method, the qubit manipulation takes place at the qubit frequency, which allows time-domain experiments even when the coherence time is very short. Its drawbacks are its lack of versatility, and the extremely short pulse rise-time necessary to reach the non-adiabatic regime.

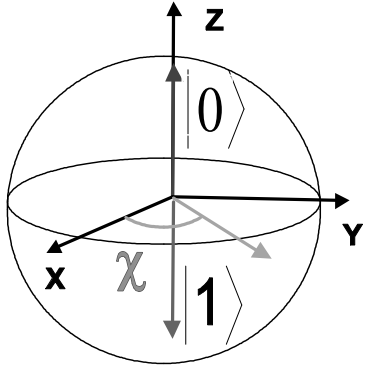

Fig. 12. Bloch sphere in the rotating frame. On resonance, the ac excitation corresponds to a static magnetic field in the equatorial plane for the fictitious spin representing the qubit (thin arrow). The angle between the field and the $x$ axis is the phase $\chi$ of the excitation with respect to the reference phase that determines the $x$ axis of the Bloch sphere.

In the second method, a control parameter is varied sinusoidally with a frequency matching the resonance requency of the qubit. This method is more versatile and more accurate than the dc pulse method, but is slower. When the gate voltage of a Cooper pair box is modulated by a resonant microwave pulse with amplitude $\delta N_G$, the Hamiltonian 3.2 contains a term $h(t) = -2E_C \left\langle 0 \left| \widehat{N} \right| 1 \right\rangle \sigma_X$, which induces Rabi precession at frequency $\omega_R = 2E_C \, \delta N_G \, /\hbar \left| \left\langle 0 \left| \widehat{N} \right| 1 \right\rangle \right|$, as shown in Fig. 12. The fictitious spin representing the qubit rotates around an axis located in the equatorial plane of the Bloch sphere. The position of this axis is defined by the relative phase $\chi$ of the microwave with respect to the microwave carrier that defines the $X$ axis. A single resonant pulse with duration $\tau$ induces a rotation by an angle $\omega_R \tau$, which manifests itself by oscillations of the switching probability, as shown in Fig. 11. When the pulse is not resonant, the detuning adds a $z$ component to the rotation vector.

### 5.2. NMR-like control of a qubit

Rabi precession, which is the basic coherent control operation, has been demonstrated for several Josephson qubits [19, 22, 24, 29]. More complex manipulations inspired from NMR [40, 41, 43] have also been applied in order to perform arbitrary single qubit gates, and to probe decoherence processes [44, 45].

Although it is possible to rotate around an out of plane axis by detuning the microwave, it is more convenient to combine on-resonance pulses. Indeed, three sequential rotations around two orthogonal axes, for instance the $x$ and $y$ axes on the Bloch sphere, should allow to perform any desired rotation. It is thus important to test whether or not two subsequent rotations combine as predicted. The result is shown in Fig. 13. A two pulse sequence was also used to probe rotations around the $z$ axis and was performed using adiabatic pulses applied to the gate charge or to the phase port. Indeed, varying the qubit frequency during a short time results in an extra phase factor between the two components of the qubit, which is equivalent to a rotation around the $Z$ axis by an angle $\varsigma = \int \delta\omega_{01}(t)dt$. As discussed further below, the two pulse sequence also probes decoherence during the free evolution of the qubit between the two pulses.

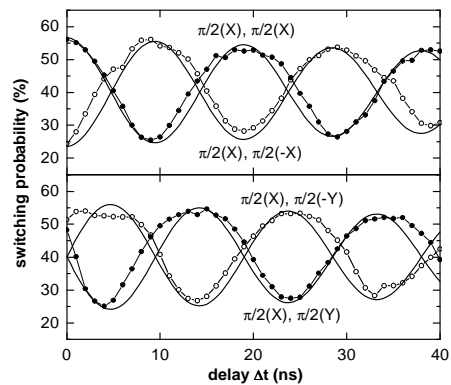

Fig. 13. Switching probability after two $\pi/2$ pulses around $x$, $y$, $-x$, or $-y$ axes, as a function of the delay between the pulses. The phase of the oscillating signal at the detuning frequency 50 MHz depends as predicted for the different combinations of rotation axes. The solid lines are theoretical fits *(taken from [45])*.

The issue of gate robustness is also extremely important because the needs of quantum computing are extremely demanding. In NMR, composite pulse methods have been developed in order to make transformations less sensitive to pulse imperfections [41–43]. In these methods, a single pulse is replaced by a series of pulses that yield the same operation, but with a decreased sensitivity to pulse imperfections. In the case of frequency detuning, a particular sequence named CORPSE (Compensation for Off-Resonance with a Pulse Sequence) has proved to be extremely efficient [42]. The sensitivity to detuning is indeed strongly re-duced, the error starting at fourth order in detuning instead of second order for a single pulse. This sequence has been probed in the case of a $\pi$ rotation around the $X$ axis. As shown in Fig. 14, it is significantly more robust against detuning than a single $\pi$ pulse. This robustness was probed starting from state 0, but also from any state with a representative vector in the YZ plane (see inset).

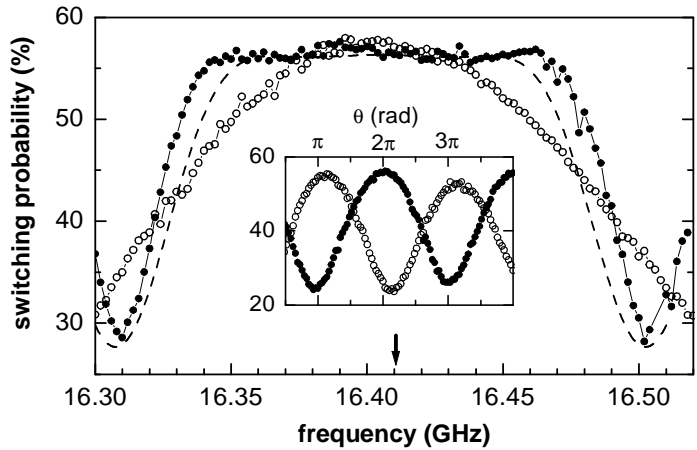

Fig. 14. Demonstration of the robustness of a composite pulse with respect to frequency detuning: switching probability after a $CORPSE$ $\pi(X)$ sequence (disks), and after a single $\pi(X)$ pulse (circles). The dashed line is the prediction for the $CORPSE$ $\pi(X)$ sequence, the arrow indicates the qubit transition frequency. The $CORPSE$ sequence works over a larger frequency range. The Rabi frequency was 92 MHz. Inset: oscillations of the switching probability after a single pulse $\theta(-X)$ followed (disks) or not (circles) by a $CORPSE$ $\pi(X)$ pulse. The patterns are phase shifted by $\pi$, which shows that the $CORPSE$ sequence is indeed equivalent to a $\pi$ pulse *(Taken from [45])*.

## 6. Probing qubit coherence

We discuss now decoherence during the free evolution of the qubit, which induces the decay of the qubit density matrix. As explained in section 3.2.3, decoherence is characterised by relaxation, affecting the diagonal and off diagonal parts of the density matrix, and by dephasing, which affects only its off diagonal part.

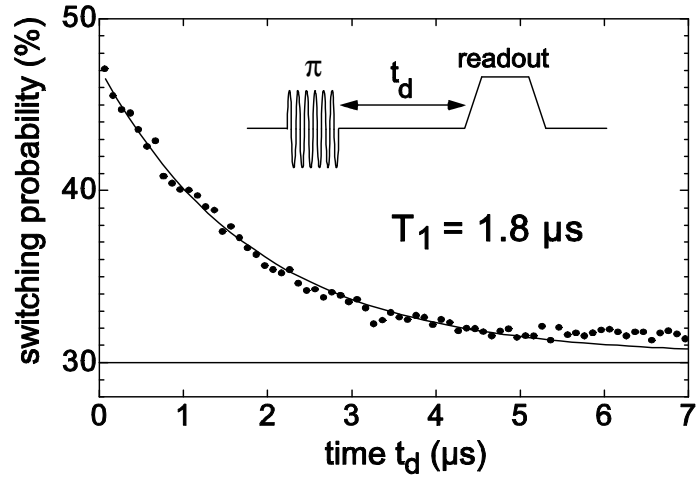

Fig. 15. Decay of the switching probability of the quantronium's readout junction as a function of the delay $t_d$ between a $\pi$ pulse that prepares state $|1\rangle$ and the readout pulse.*(Taken from [24])*

### 6.1. Relaxation

Relaxation is readily obtained from the decay of the signal after a $\pi$ pulse, as shown in Fig. 15. The relaxation time in the quantronium ranges from a few hundreds of nanoseconds up to a few microseconds. These relaxation times are shorter than those calculated from the coupling to the external circuit using an estimated value for the asymmetry factor $d$. Note however that the electromagnetic properties of the circuit are difficult to evaluate at the qubit transition frequency. Since a similar discrepancy is found in all Josephson qubits, this suggests that qubits with a simple microwave design are preferable, and that microscopic relaxation channels may be present in all these circuits, as suggested by recent experiments on phase qubits [46]. A confirmation of this would imply the necessity of a better junction technology.

### 6.2. Decoherence during free evolution

The most direct way to probe decoherence is to perform a Ramsey fringe experiment, as shown in Fig. 16, using two $\pi/2$ pulses slightly out of resonance. The first pulse creates a superposition of states, with an off diagonal density matrix. After a period of free evolution, during which decoherence takes place, a second pulse transforms the off-diagonal part of the density matrix into a longitudinal component, which is measured by the subsequent readout pulse. The decay of the obtained oscillations at the detuning frequency characterise decoherence. This experiment was first performed in atomic physics, and it corresponds to the free induction decay (FID) in NMR. When the decay is not exponential, we define the coherence time as the time corresponding to a decay factor $\exp(-1)$. Other more sophisticated pulse methods have been developed to probe coherence. When the operating point is moved away from the optimal point at which decoherence is weak during a fraction of the delay between the two pulses of a Ramsey sequence, the signal gives access to decoherence at this new working point. The interest of this 'detuning' method is to perform qubit manipulations at the optimal working point without being hindered by strong decoherence. When the coherence time is too short for time domain experiments, the lineshape, which is the Fourier transform of the Ramsey signal, gives access to the coherence time. Coherence times obtained with all these methods on a single sample away from the optimal point in the charge and phase directions are indicated by full symbols in Fig. 17.

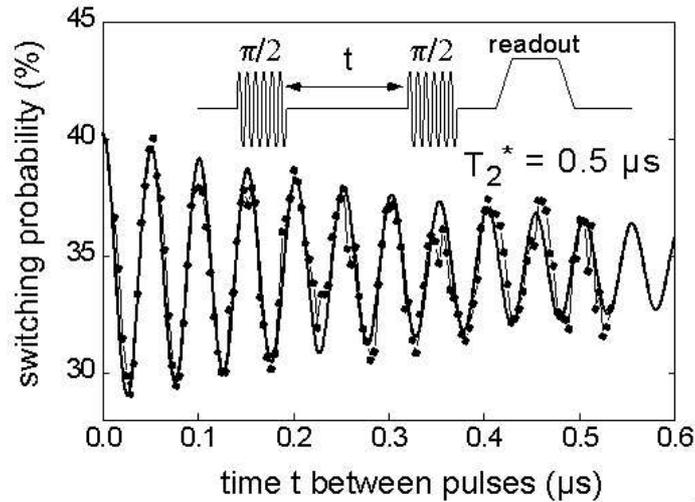

Fig. 16. Ramsey fringe experiment on a quantronium sample at the optimal point. Two $\pi/2$ microwave pulses slightly out of resonance and separated by a time delay $t$ are applied to the gate, The oscillations of the switching probability (dots) at the detuning frequency give direct access to decoherence. In this experiment, their decay time was 500 $ns$, as estimated by the fit to an exponentially decaying cosine (full line). Coherence times have been measured to be in the range $200 - 500 \ ns$ for the quantronium *(Quantronics group).*

It is possible to shed further light on the decoherence processes and to fight them using the echo technique well known in NMR [40]. An echo sequence is a two $\pi/2$ pulse Ramsey sequence with a $\pi$ pulse in the middle, which causes the phase accumulated during the second half to be subtracted from the phase accumulated during the first half. When the noise-source producing the frequency fluctuation is static on the time scale of the pulse sequence, the echo does not decay. The observed echo decay times, indicated by open disks in Fig. 17, thus set constraints on the spectral density of the noise sources. In particular, these data indicate that the charge noise is significantly smaller than expected from the low frequency $1/f$ spectrum, at least in the two samples in which echo experiments were performed. Bang-bang suppression of dephasing, which generalises the echo technique, could fight decoherence more efficiently [47].

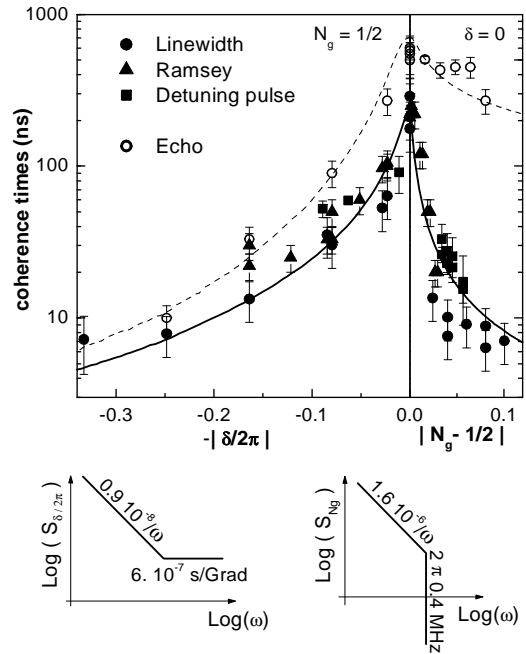

Fig. 17. Coherence times $T_2$ and $T_{Echo}$ in a quantronium sample extracted from the decay of free evolution signals. The full and dashed lines are calculated using the the spectral densities depicted by the bottom graphs for the phase noise (left) and for the charge noise (right), respectively. *(Quantronics group)*.

*6.3. Decoherence during driven evolution*

During driven evolution, the density matrix is best defined using the eigenstate basis in the rotating frame. On resonance, these eigenstates are the states $|X\rangle$ and $|-X\rangle$ on the Bloch sphere. As in the laboratory frame, the decay of the density matrix involves relaxation and dephasing. The measurement of the relaxation time can be performed using the so-called spin locking technique in NMR [40], which allows one to measure the qubit polarisation after the preparation of the state $|X\rangle$. The coherence time during driven evolution is easily obtained from Rabi oscillations. Indeed, the initial state $|0\rangle$ is a coherent superposition of the eigenstates during driven evolution since $|0\rangle = (|X\rangle + |-X\rangle) / \sqrt{2}$. The Rabi signal measured after a pulse of duration $t$ thus probes decoherence during driven evolution. The corresponding coherence time is longer than the coherence time during free evolution because the driving field quenches the effect of the low frequency fluctuations that dominate dephasing during free evolution.

## 7. Qubit coupling schemes

Single qubit control and readout has been achieved for several Josephson qubits. Although the control accuracy and readout fidelity do not yet meet the requirements for quantum computing, the demonstration on such 'working' qubits of logic gates is now the main goal. Presently, only a few experiments have been performed on coupled qubits. A logic $C - NOT$ gate was operated in 2003 on charge qubits [48], but without a single shot readout. The correlations between coupled phase qubits have been measured recently using a single-shot readout [49]. However, the entanglement between two coupled qubits has not yet been investigated with sufficient accuracy to probe the violation of Bell inequalities predicted by quantum mechanics. Only such an experiment can indeed test if collective degrees of freedom obey quantum mechanics, and whether or not the entanglement decays as predicted from the known decoherence processes.

*7.1. Tunable versus fixed couplings*

In a processor, single qubit operations have to be supplemented with two qubit logic gate operations. During a logic gate operation, the coupling between the two qubits has to be controlled with great accuracy. For most solid state qubits, there is however no simple way to switch on and off the coupling and to control its amplitude. In the cases of the implementations based on P impurities in silicon and on electrons in quantum dots, the exchange energy between two electrons, which can be varied with a gate voltage, provides a tunable coupling. In the case of the superconducting qubits, controllable coupling circuits have been proposed, but fixed coupling Hamiltonians have been mostly considered: capacitive coupling for phase, charge-phase and charge qubits, and inductive coupling for flux qubits. It is nevertheless possible to use a constant coupling Hamiltonian provided that the effective qubit-qubit interaction induced by this coupling Hamiltonian is controlled by other parameters. We now discuss all these coupling schemes.

### 7.2. A tunable coupling element for Josephson qubits

The simplest way to control the coupling between two Josephson qubits is to use a Josephson junction as a tunable inductance. For small phase excursions, a Josephson junction with phase difference $\delta$ behaves as an effective inductance $L = \varphi_0 / (I_0 \cos \delta)$. Two Josephson junctions in parallel form an effective junction whose inductance can be controlled by the magnetic flux through the loop. When an inductance is placed in a branch shared by two qubit loop circuits, which is possible for phase, charge-phase and flux qubits, the coupling between the two qubits is proportional to the branch inductance. Note that, in this tunable coupling scheme, the qubits have to be moved slightly away from their optimal working point, which deteriorates quantum coherence. The spectroscopy of two flux qubits whose loops share a common junction has been performed [51], and been found to be in close agreement with the predictions. In the case of two charge-phase qubits sharing a junction in a common branch [52], the coupling takes a longitudinal form in the qubit eigenstate basis: $H_{cc} = -\hbar\omega_C \, \widehat{\sigma}_{Z1}\widehat{\sigma}_{Z2}$, where $\omega_C$ is the coupling frequency. The amplitude of the effective $z$ field acting on each fictitious spin is changed proportionally to the $z$ component of the other spin. This coupling allows one to control the phase of each qubit, conditional upon the state of the other one.It thus allows the implementation of the Controlled Phase gate, from which the controlled not $CNOT$ gate can be obtained.

### 7.3. Fixed coupling Hamiltonian

The first demonstration of a logic gate was performed using a fixed Hamiltonian. The system used consisted of two Cooper pair boxes with their islands connected by a capacitance $C_C$. The coupling Hamiltonian is

$$H_{cc} = -E_{CC}(\widehat{N}_1 - N_{G1})(\widehat{N}_2 - N_{G2}) \tag{7.1}$$

where $E_{CC} = -E_{C1}E_{C2}C_C/(2e)^2$ is the coupling energy, smaller than the charging energy of the Cooper pair boxes. This Hamiltonian corresponds to changing the gate charges by $(E_{CC}/2E_{C1})/(\widehat{N}_2 - N_{G2})$ for qubit 1, and by $(E_{CC}/2E_{C2})/(\widehat{N}_1 - N_{G1})$ for qubit 2. The correlations between the two qubits predicted for this Hamiltonian have been probed, as shown in Fig. 18. A C-NOT logic gate was operated with this circuit [48].

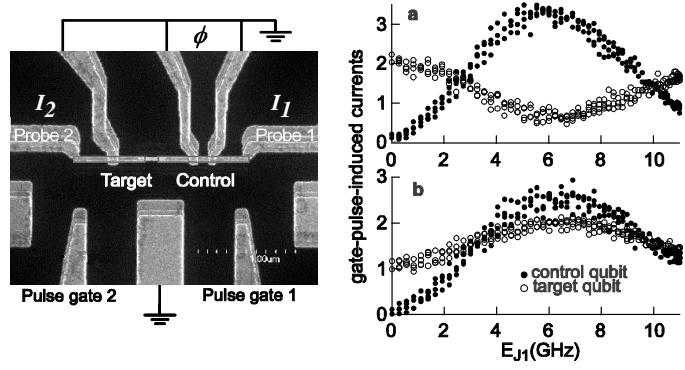

Fig. 18. Demonstration of the correlation between two capacitively coupled charge qubits. Pulse-induced current as a function of the Josephson energy of the control qubit. The control qubit is prepared in a superposition of states that depends on its Josephson energy $E_{J1}$. A pulse applied on the target qubit yields a $\pi$ rotation only when the control qubit is in state $|0\rangle$. The currents through the two probe junctions can be anticorrelated (a) or correlated (b) when $E_{J1}$ is varied. *(Courtesy of T. Yamamoto et al. [48], NEC, Japan).*

In the uncoupled eigenstate basis, The Hamiltonian (7.1) contains both longitudinal terms of type $\hat{\sigma}_{Z1}\hat{\sigma}_{Z2}$ and transverse terms of type $\hat{\sigma}_{X1}\hat{\sigma}_{X2}$. At the double optimal point $N_{G1} = N_{G2} = 1/2$, $\delta_1 = \delta_2 = 0$, the Hamiltonian (7.1) is transverse $H_{CC} = \hbar\Omega_C\hat{\sigma}_{X1}\hat{\sigma}_{X2}$, with $\Omega_C = E_{CC}/\hbar \left|\left\langle 0_1\hat{N}_1 1_1\right\rangle\right| \left|\left\langle 0_2\left|\hat{N}_2\right|1_2\right\rangle\right|$. When the two qubits have the same resonance frequency $\omega_{01}$, and when $\Omega_C \ll \omega_{01}$, the non-secular terms in $H_{CC}$ that do not commute with the single qubit Hamiltonian are ineffective, and the effective Hamiltonian reduces to:

$$H_{CC}^{\mathrm{sec}} = (\hbar\Omega_C) \; (\hat{\sigma}_{+1}\hat{\sigma}_{-2} + \hat{\sigma}_{-1}\hat{\sigma}_{+2}) \,. \tag{7.2}$$

The evolution of the two qubits corresponds to swapping them periodically. More precisely, a swap operation is obtained at time $\pi/\Omega_C$. This gate is called $ISWAP$ because of extra factors $i$:

$ISWAP\,|00\rangle = |00\rangle$ ; $ISWAP\,|10\rangle = -i\,|01\rangle$ ;

$ISWAP\,|01\rangle = -i\,|10\rangle$ ; $ISWAP\,|11\rangle = |11\rangle$.

At time $\pi/4\Omega_C$, the evolution operator corresponds to the gate $\sqrt{ISWAP}$, which is universal.

### 7.4. Control of the interaction mediated by a fixed Hamiltonian

The control of the qubit-qubit interaction mediated by a fixed Hamiltonian depends on the form of this Hamiltonian.

For a coupling of the form 7.2, the effective interaction can be controlled by varying the qubit frequencies since the qubits are affected only when their frequency difference is smaller than $\Omega_C$. This tuning strategy was recently applied to capacitively coupled phase qubits, in which the qubit frequency is directly controlled by the bias current of the junctions [6]. The correlations predicted by quantum mechanics between the readouts of the two qubits were observed [49]. The tuning strategy would be also well suited for coupling many qubits together through an oscillator [28]. The virtual exchange of photons between each qubit and the oscillator indeed yields a coupling of the form 7.2, which is efficient only when the two qubits are tuned. This coupling scheme yields truly scalable designs, whereas most of other schemes are limited to 1D qubit arrays, with nearest neighbor couplings. The coupling between a qubit and a resonator has been already demonstrated for the charge and flux qubits [32, 50].

Another method proposed recently consists in maintaining the qubits out of resonance, but in reaching an equivalent resonance condition in the presence of resonant microwave pulses applied to each one [53]. This method is based on an NMR protocol developed by Hartmann and Hahn in order to place two different spin species 'on speaking terms'. In this scheme, the energy difference between the two qubits is exchanged with the microwave fields.

The case of the longitudinal coupling $H_{CC} \propto -\hbar\omega_C\widehat{\sigma}_{Z1}\widehat{\sigma}_{Z2}$ has not been considered yet. Although the control of this coupling is commonly performed in high resolution NMR, adaptation to qubits has not been attempted.

### 7.5. Running a simple quantum algorithm

Despite the fact that no quantum processor is yet available, running a simple quantum algorithm in a Josephson qubit circuit is nonetheless presently within reach. Let us consider Grover's search algorithm, which is able to retrieve an object among $N$ in $\sqrt{N}$ algorithmic steps [1]. In the simple case of 4 objects, it requires a single algorithmic step. Let us consider a two-qubit system $\{1, 2\}$ with an $ISWAP$ gate. The object to be retrieved is an operator $\widehat{O}$ taken among the four operators $R_{1Z}(\pm\pi/2)R_{2Z}(\pm\pi/2)$, where $R_U(\alpha)$ denotes a rotation around the $u$ axis by an angle $\alpha$. A simplified version of Grover's algorithm proceeds as follows:

-first, a superposition of all eigenstates is prepared by applying single qubit rotations around the $y$ axis:

$$|\Psi\rangle = 1/2(|00\rangle + |10\rangle + |01\rangle + |11\rangle)$$

We assume that single qubit rotations are fast enough to neglect the effect of the two qubit interaction during their duration.

-A single algorithm step is then applied, with the operator:

$$U = R_{1X}(\pi/2)R_{2X}(\pi/2) \; ISWAP \; \widehat{O} \; ISWAP.$$

-The state $U \left| \Psi \right\rangle$ is then read, and the outcome determines which operator had been selected. For instance, the outcome $\left| 11 \right\rangle$ corresponds to the operator $\widehat{O} = R_{1Z}(\pi/2)R_{2Z}(\pi/2)$.

With more qubits, more sophisticated quantum manipulations and algorithms become possible. Note in particular that teleportation is possible with 3 qubits [1].

## 8. Conclusions and perspectives

Many solid state qubits have been proposed, and several of them have already demonstrated coherent evolution.

For semiconductor qubits, the coherent transfer of an electron between two dots has been demonstrated, and other promising designs are under investigation.

For superconducting qubits, single qubit control, single-shot readout, and a two-qubit logic gate have been achieved. Methods inspired from NMR have been applied to qubit manipulation in order to improve its robustness, and to probe decoherence processes. The lack of an efficient readout scheme and of robust two qubit gates still hinders the development of the field. New QND readout schemes are presently investigated in order to reach a higher readout fidelity. Different two qubit gates have been proposed, but none of them is as robust as the *NAND* gate used in ordinary classical processors. Currently, the coherence time, the readout fidelity, and the gate accuracy are insufficient to envision quantum computing. But how far from this goal are solid state qubits?

In order to use quantum error correcting codes, an error rate of the order of $10^{-4}$ for each logic gate operation is required. Presently, the gate error rates can be estimated at about a few % for single qubit gates, and at about 20% at best for two qubit gates. The present solid state qubits thus miss the goal by many orders of magnitude. When decoherence and readout errors are taken into account, quantum computing appears even more unrealistic. This is not, however, a reason to give up because conceptual and technical breakthroughs can be expected in this rather new field, and because no fundamental objection has been found. One should not forget that, in physics, everything which is possible is eventually done. Furthermore, quantum circuits provide new research directions in which fundamental questions on quantum mechanics can be addressed. The extension of quantum entanglement out of the microscopic world, and the location and nature of the frontier between quantum and classical worlds, are two of these essential issues. For instance, the accurate measurement of the correlations between two coupled qubits would indeed probe whether or not the collective variables of qubit circuits do follow quantum mechanics.

Our feeling is that, whatever the motivation, complex quantum systems and quantum machines are a fascinating field worth the effort.

*Acknowledgements:*   The qubit research is a collective effort carried out by groups worldwide. We acknowledge discussions with many colleagues from all these groups. We thank in particular the participants to the european project SQUBIT, whose input has been instrumental. The research on the Quantronium has been carried in the Quantronics group at CEA-Saclay. We warmly thank all the group members and all the visitors for maintaining a demanding but friendly research atmosphere. We thank P. Meeson and N. Boulant for their help with the manuscript. We acknowledge the support  from the CEA, the CNRS, and of the european project SQUBIT. Last but not least, we thank all those who contribute to make the Les Houches school a so lively place.  Teaching there is a unique experience.

# References

[1]  M.A. Nielsen and I.L. Chuang, "*Quantum Computation and Quantum Information*" (Cambridge University Press, Cambridge, 2000.

[2]  *Quantum Coherence and Information Processing* , edited by D. Esteve, J.M. Raimond, and J. Dalibard (Elsevier, 2004).

[3]  I. Chuang, course 1 in ref. 2.

[4]  C. Glattli, course 11 in ref. 2.

[5]  M.H. Devoret and J. Martinis, course 12 in ref. 2.

[6]  J. Martinis, course 13 in ref. 2.

[7]  D. Vion, course 14 in ref. 2.

[8]  M.H. Devoret , D.Esteve ,C. Urbina , J.M. Martinis, A.N. C leland, and J.Clarke, in "*Quantum Tunneling in Condensed Media*", Kagan N Yu., Leggett A.J., eds. (Elsevier Science Publishers, 1992) pp. 313-345.

[9]  A. Aspect, P. Grangier, and Gérard Roger, Phys. Rev. Lett **49**, 91 (1982).

[10]  S. Haroche, course 2 in ref. 2; M. Brune, course 3 in ref. 2.

[11]  R. Blatt, H. Häffner, C.F. Ross, C. Becher, and F. Schmidt-Kaler, course 5 in ref. 2; D.J. Wineland, course 6 in ref. 2.

[12]  A. Zazunov, V. S. Shumeiko, E. N. Bratus', J. Lantz, and G. Wendin, Phys. Rev. Lett. 90, 087003 (2003).

[13]  B. E. Kane, Nature.393, 133 (1998).

[14]  J. M. Elzerman, R. Hanson, L. H. Willems van Beveren, B. Witkamp, J. S. Greidanus, R. N. Schouten, S. De Franceschi, S. Tarucha, L. M. K. Vandersypen, and L.P. Kouwenhoven, *Quantum Dots: a Doorway to Nanoscale Physics*, in *Series: Lecture Notes in Physics*, **667**, Heiss, WD. (Ed.), (2005), and refs. therein.

[15]  T. Hayashi, T. Fujisawa, H. D. Cheong, Y. H. Jeong, and Y. Hirayama, Phys. Rev. Lett. 91, 226804 (2003).

[16]  A. Barone and G. Paternò, *Physics and applications of the Josephson effect* (Wiley, New York, 1982).

[17]  M.H. Devoret, in "*Quantum Fluctuations*", S. Reynaud, E. Giacobino, J. Zinn-Justin, eds. (Elsevier, Amsterdam, 1996), p.351.

[18]  Guido Burkard, Roger H. Koch, and David P. DiVincenzo, Phys. Rev. B 69, 064503 (2004).

[19]  Y. Nakamura, Yu. A. Pashkin and J. S. Tsai, Nature **398**, 786, (1999).

[20]  J. M. Martinis, S. Nam, J. Aumentado, and C. Urbina, Phys. Rev. Lett. 89, 117901 (2002).

[21]  J. E. Mooij, T. P. Orlando, L. Levitov, Lin Tian, Caspar H. van der Wal, and Seth Lloyd, Science 285, 1036 (1999).

[22]  I. Chiorescu, Y. Nakamura, C. J. P. M. Harmans, and J. E. Mooij, Science 299, 1869 (2003).

[23]  A. Cottet, D. Vion, P. Joyez, P. Aassime, D. Esteve, and M.H. Devoret, Physica C **367**, 197 (2002).

[24]  D. Vion *et al.*, Science **296**, 886 (2002).

[25]  A. Cottet, *Implementation of a quantum bit in a superconducting circuit*, PhD thesis, Université Paris VI, (2002); www-drecam.cea.fr/drecam/spec/Pres/Quantro/ .

[26]  V. Bouchiat, D. Vion, P. Joyez, D. Esteve and M.H. Devoret, Physica Scripta, **76**, 165 (1998); V. Bouchiat, PhD thesis, Université Paris VI, (1997), www-drecam.cea.fr/drecam/spec/Pres/Quantro/ .

[27]  *Single Charge Tunneling*, edited by H. Grabert and M. H. Devoret (Plenum Press, New York, 1992).

[28]  Y. Makhlin, G. Schön and A. Shnirman, Rev. Mod. Phy, **73**, 357 (2001).

[29]  T. Duty, D. Gunnarsson, K. Bladh, and P. Delsing, Phys. Rev. **B 69**, 140503 (2004).

[30]  R.J. Schoelkopf *et al.*, Science, **280**, 1238 (1998).

[31]  O. Astafiev, Yu. A. Pashkin, Y. Nakamura, T. Yamamoto, and J. S. Tsai, Phys. Rev. Lett. **93**, 267007 (2004).

[32]  A. Wallraff, D. Schuster,.-I.; A. Blais; L. Frunzio; R.-S. Huang,- J. Majer, S. Kumar, S.M.Girvin, R.J. Schoelkopf, Nature **431**, 162 (2004); and p. 591 in ref. 2.

[33]  E. Paladino, L. Faoro, G. Falci, and Rosario Fazio, Phys. Rev. Lett. 88, 228304 (2002).

[34]  Y. Makhlin and A. Shnirman, Phys. Rev. Lett. 92, 178301 (2004).

[35]  A. Lupascu, .J.M.Verwijs, R.N. Schouten, C.J.P.M. Harmans, and J.E. Mooij, Phys. Rev. Lett. **93**, 177006 (2004).

[36]  I. Siddiqi, R. Vijay, F. Pierre, C. M. Wilson, M. Metcalfe, C. Rigetti, L. Frunzio,R.J. Schoelkopf, M. H. Devoret, D. Vion, and D. Esteve, Phys. Rev. Lett. **94**, 027005 (2005).

[37]  I. Siddiqi, R. Vijay, F. Pierre, C. M. Wilson, M. Metcalfe, C. Rigetti, L. Frunzio, and M. H. Devoret, Phys. Rev. Lett. **93**, 207002 (2004).

[38]  Mika A. Sillanpää, Leif Roschier, and Pertti J. Hakonen, Phys. Rev. Lett. **93**, 066805 (2004).

[39]  E. Fahri, J. Goldstone, S. Gutmann, and M. Sipser. Science **292**, 472 (2001).

[40]  C.P. Slichter, *Principles of Magnetic Resonance*, Springer-Verlag (3rd ed: 1990).

[41]  J. Jones, course 10 in ref. 2.

[42] H.K. Cummins, G. Llewellyn, and J.A. Jones, Phys. Rev. A **67**, 042308 (2003).

[43] L.M.K. Vandersypen and I.L. Chuang, quant-ph/0404064.

[44] D. Vion *et al.*, Fortschritte der Physik, **51**, 462 (2003).

[45] Collin E., Ithier G., Aassime A., Joyez P., Vion D., Esteve D. Phys. Rev. Lett. **93**, 157005 (2004).

[46] K. B. Cooper, Matthias Steffen, R. McDermott, R. W. Simmonds, Seongshik Oh, D. A. Hite, D. P. Pappas, and John M. Martinis, Phys. Rev. Lett. **93**, 180401 (2004).

[47] G. Falci, A. D'Arrigo, A. Mastellone, and E. PaladinoPhys. Rev. A **70**, 040101 (2004); H. Gutmann, F.K. Wilhelm, W.M. Kaminsky, and S. Lloyd, Quantum Information Processing, Vol. **3**, 247 (2004).

[48] T. Yamamoto et al., Nature **425**, 941 (2003), and Yu. Pashkin *et al.*, Nature **421**, 823 (2003).

[49] R. McDermott, R. W. Simmonds, Matthias Steffen, K. B. Cooper, K. Cicak, K. D. Osborn, Seongshik Oh, D. P. Pappas, and John M. Martinis, Science **307**, 1299 (2005).

[50] I. Chiorescu, P. Bertet, K. Semba, Y. Nakamura, C. J. P. M. Harmans, and J. E. Mooij, Nature **431**, 159 (2004).

[51] Hans Mooij, private communication.

[52] J. Q. You, Y. Nakamura, and F. Nori, Phys. Rev. B 71, 024532 (2005); J. Lantz, M. Wallquist, V. S. Shumeiko, and G. Wendin, Phys. Rev. B **70**, 140507 (2004).

[53] C. Rigetti and M.H. Devoret, quant-ph/0412009.